\documentclass[tightenlines,floats,aps,nofootinbib,prd,onecolumn,preprintnumbers]{revtex4}

\usepackage[dvipdfmx]{graphicx}
\usepackage[caption=false]{subfig}
\usepackage{amsmath,amssymb,algorithm,algorithmic,bm,color}

\allowdisplaybreaks[1]

\newcommand{\appname}{`{\tt Spectrum Reconstructor}'~}

\newcommand{\PP}{\mathcal{P}}
\newcommand{\calR}{\mathcal{R}}

\begin{document}

\title{Reconstruction of primordial tensor power spectra from B-mode
polarization of the cosmic microwave background}
\author{
Takashi Hiramatsu$^1$,
Eiichiro Komatsu$^{2,3}$,
Masashi Hazumi$^{3,4,5,6}$,
Misao Sasaki$^{7,8}$
}

\affiliation{
$^1$ Department of Physics, Rikkyo University, Toshima, Tokyo, 171-8501, Japan\\
$^2$ Max-Planck-Institut f\"ur Astrophysik, Karl-Schwarzschild Strasse 
1, 85748 Garching, Germany \\
$^3$ Kavli Institute for the Physics and Mathematics of the Universe
(Kavli IPMU, WPI), Todai Institutes for Advanced Study, the University of
Tokyo, Kashiwa 277-8583, Japan\\
$^4$High Energy Accelerator Research Organization (KEK), Tsukuba,
Ibaraki 305-0801, Japan\\
$^5$ SOKENDAI (The Graduate University for Advanced Studies), Hayama, Miura
District, Kanagawa 240-0115, Japan\\
$^6$ Institute of Space and Astronautical Science (ISAS), Japan Aerospace
Exploration Agency (JAXA), Sagamihara, Kanagawa 252-0222, Japan\\
$^7$ Center for Gravitational Physics, Yukawa Institute for Theoretical Physics, Kyoto University, Kyoto 606-8502, Japan\\
$^8$ International Research Unit of Advanced Future Studies, Kyoto University, Kyoto 606-8502, Japan
}
\begin{abstract}
Given observations of B-mode polarization power spectrum of the cosmic
 microwave background (CMB), we can reconstruct power spectra of
 primordial tensor modes from the early Universe without assuming their
 functional form such as a power-law spectrum. Shape of the reconstructed
 spectra can then be used to probe the origin of tensor modes in a
 model-independent manner. We use the Fisher matrix to calculate the
 covariance matrix of tensor power spectra reconstructed in bins.
 We find that the power spectra are best reconstructed at
 wavenumbers in the vicinity of $k\approx 6\times 10^{-4}$ and $5\times
 10^{-3}~{\rm Mpc}^{-1}$, which correspond to the ``reionization bump''
 at $\ell\lesssim 6$ and ``recombination bump'' at $\ell\approx 80$ of
 the CMB B-mode power spectrum, respectively. The error bar between
 these two wavenumbers is larger because of lack of the signal between
 the reionization and recombination bumps. The error bars increase
 sharply towards smaller (larger) wavenumbers because of the cosmic
 variance (CMB lensing and instrumental noise).
 To demonstrate utility of the reconstructed power spectra we
 investigate whether we can distinguish between various sources of tensor
 modes including those from the vacuum metric fluctuation and SU(2) gauge
 fields during single-field slow-roll inflation, open inflation and
 massive gravity inflation.  The results depend on the
 model parameters, but we find that future CMB experiments are sensitive
 to differences in these models. We make our calculation tool available
 on-line.
\end{abstract}

\preprint{RUP-18-7, YITP-18-13}

\maketitle

\section{Introduction}
Primordial gravitational waves from the very early Universe
generate B-mode polarization in the cosmic microwave background (CMB)
\cite{Seljak:1996gy,Kamionkowski:1996zd}. Usually,
we calculate the angular power spectrum of B-mode polarization by
assuming a specific form (e.g., a power law) of the power spectrum of
gravitational waves (tensor perturbations) in the early Universe and numerically
evolving tensor perturbations forward with a linear Boltzmann code such
as CMBFAST\footnote{\tt
https://lambda.gsfc.nasa.gov/toolbox/tb\_cmbfast\_ov.cfm}
\cite{Seljak:1996is}, CAMB\footnote{\tt https://camb.info/}
\cite{Lewis:1999bs}, and CLASS\footnote{\tt http://class-code.net/} \cite{Blas:2011rf}.

It is also possible to reconstruct initial tensor power spectra in
bins of wavenumbers from an observed CMB B-mode power spectrum. This is
possible when the transfer function that relates the initial
(primordial) tensor power to that at late times depends only on
the standard cosmological parameters, and not on the nature of initial
tensor perturbations. In this paper we use inflation
\cite{Brout:1977ix,Starobinsky:1980te,Sato:1980yn,Guth:1980zm,Albrecht:1982wi,Linde:1981mu}
as an example.

Inflation can produce primordial tensor perturbations
from either the vacuum fluctuation in metric \cite{Starobinsky:1979ty} or matter
fields (see e.g., \cite{Dimastrogiovanni:2016fuu} and references
therein). The vacuum metric fluctuation in single-field slow-roll inflation
models typically yields a nearly scale-invariant tensor power spectrum
\cite{Abbott:1984fp}, whereas the sourced
tensor modes can be strongly scale-dependent
\cite{Dimastrogiovanni:2016fuu}. In addition, tensor perturbations from
open inflation \cite{Tanaka:1997kq} 
and massive gravity inflation (see e.g., \cite{Domenech:2017kno} and 
references therein, and also see Appendix \ref{appsub:model}) can 
produce scale-dependent tensor perturbations. 
It is always possible to test these models individually by assuming
a functional form of the initial tensor power spectrum, evolving it
forward, and comparing to the observed B-mode power spectrum; however,
reconstructing the tensor power spectrum from the observed B-mode power
spectrum allows us to directly test various sources of the tensor
perturbation. In addition, as the reconstruction does not depend on the
nature of initial tensor perturbations, it may reveal unexpected
features in the initial tensor power spectrum in a model-independent
manner. In this paper, we demonstrate this point using the Fisher matrix
formalism.

The rest of the paper is organized as follows. In Sec.~\ref{sec:method}
we describe our methodology. In Sec.~\ref{sec:results} we obtain the
covariance matrix of the reconstructed tensor power spectrum and show
how to distinguish between various models. We conclude in
Sec.~\ref{sec:conclusion}.

\section{Methodology}
\label{sec:method}
We parameterize the primordial tensor power spectrum by $N$ bins in
logarithmic intervals,
%
\begin{equation}
 \PP_h(k) = 
\begin{cases}
\PP_h^{\rm fid}(k) + \delta \PP_i & {\rm for}\;k_{i-1}\leq k<k_i\;{\rm
 with}\;1\leq i\leq N\,, \\
\PP_h^{\rm fid}(k)              & {\rm for}\; k<k_0\;{\rm and}\;k_N\leq k\,,
 \end{cases}
 \label{eq:bandpower}
\end{equation}
%
where $\PP_h(k) = (k^3/2\pi^2)P_h(k)$ is the dimensionless amplitude
of the tensor power spectrum, $\delta \PP_i$'s are constants, and
$k_n=\alpha^nk_0$ with a constant $\alpha$ controlling the logarithmic
interval. In this paper, we shall take a power-law spectrum as the fiducial
power spectrum $\PP_h^{\rm fid}(k)$:
%
\begin{equation}
\PP_h^{\rm fid}(k) = r\PP_{\calR 0}\left(\frac{k}{k_{\rm pivot}}\right)^{n_T},
\label{eq:fiducial}
\end{equation}
%
where $r$ is the tensor-to-scalar ratio and $\PP_{\calR 0}$ is the
amplitude of curvature perturbations at the pivot scale, $k=k_{\rm
pivot}=0.002~{\rm Mpc}^{-1}$.

We use the Fisher matrix to compute the covariance matrix of
$\delta\PP_i$ given measurement uncertainties in the B-mode
observations. The Fisher matrix is given by
%
\begin{equation}
 F_{ij} = f_{\rm sky}\sum_{\ell=2}^{\ell_{\rm max}}\frac{2\ell+1}{2}
  \frac{1}{\mathcal{N}_\ell^2}
 \left(\frac{\partial C^{\rm BB}_\ell}{\partial\delta \PP_i}\right)
 \left(\frac{\partial C^{\rm BB}_\ell}{\partial\delta \PP_j}\right),
\label{eq:fisher}
\end{equation}
%
where $f_{\rm sky}$ is a fraction of the sky observed, and
%
\begin{equation}
 \frac{\partial C^{\rm BB}_\ell}{\partial\delta \PP_i}
 =
  4\pi\int_{k_{i-1}}^{k_i}\!T^{(T)}_{B\ell}{}^2(k)\,\frac{dk}{k},
\end{equation}
%
with the tensor B-mode transfer function $T^{(T)}_{B\ell}$.

As for the noise contributions, we use
%
\begin{equation}
  \mathcal{N}_\ell = C_\ell^{\rm BB,fid} + \lambda C_\ell^{\rm BB,lens} + N_\ell\exp(\ell^2\sigma_b^2).
 \label{eq:noise}
\end{equation}
%
Here $C_\ell^{\rm BB,fid}$ is the the angular power spectrum
of B-mode polarization from the fiducial tensor power spectrum:
%
\begin{equation}
 C_\ell^{\rm BB,fid} = 4\pi\int T_{B\ell}^{(T)2}(k)\PP_h^{\rm fid}(k)\frac{dk}{k}.
\end{equation}
%
We use {\tt cmb2nd} \cite{Hiramatsu} to compute the transfer function
with the cosmological parameters from the Planck 2015 results
(TT,TE,EE$+$lowP$+$lensing$+$ext in Ref.~\cite{Ade:2015xua}), which are
tabulated in Table \ref{tab:param}. We have checked that the results of
{\tt cmb2nd} and CAMB agree precisely.

\begin{table}[!ht]
\centering
\begin{tabular}{lll}
\hline
amplitude of curvature perturbation & $\PP_{\calR 0}$ & $2.441\times 10^{-9}$ \\
pivot scale                         & $k_{\rm pivot}$ & $0.002~{\rm Mpc}^{-1}$ \\
spectral index                      & $n_s$           & $0.9667$ \\
reduced Hubble parameter            & $h$             & $0.6774$ \\
dark matter fraction                & $h^2\Omega_{\rm CDM}$ & $0.1188$ \\
baryon fraction                     & $h^2\Omega_{\rm b}$   & $0.02230$ \\
effective number of neutrinos       & $N_{\rm eff}$         & $3.046$ \\
photon's temperature                & $T_{\gamma,0}$        & $2.7255~{\rm K}$ \\
optical depth                       & $\tau$                & 0.066 \\
Helium abundance                    & $Y_p$                 & 0.24667 \\
\hline
\end{tabular} 
 \caption{Fiducial cosmological parameters provided by Planck 2015 results
(TT,TE,EE$+$lowP$+$lensing$+$ext in Ref.~\cite{Ade:2015xua}).}
 \label{tab:param}
\end{table}

The second term in Eq.~(\ref{eq:noise}), $C_\ell^{\rm BB,lens}$, is the
contribution from CMB lensing \cite{Zaldarriaga:1998ar}. The parameter
$\lambda$ is a ``delensing factor'', being $0$ if the lensing effect
is completely removed. The lensing B-mode induced by the scalar
perturbations is given by (e.g. Ref.~\cite{Namikawa:2015tba}, and references
therein) 
%
\begin{align}
 C_\ell^{\rm BB,lens} = \frac{1}{2\ell+1}
  \sum^{\ell'_{\rm max}}_{\ell'L}(\mathcal{S}^{(-)}_{\ell\ell'L})^2C_{\ell'}^{\rm EE}C_{L}^{\phi\phi},
\end{align}
%
where $C^{\rm EE}_\ell$ is the angular power spectrum of E-mode induced
by scalar perturbations and $C^{\phi\phi}_\ell$ is that of the lensing
potential \cite{Lewis:2006fu}. 
To obtain $C_\ell^{\rm BB,lens}$ for $\ell\leq 500$ with sufficient
accuracy, we sum up the right-hand side up to $\ell'_{\rm max}=2000$.
We find that our $C_\ell^{\rm BB,lens}$ agrees with that of CAMB
to within 0.2\% accuracy at $\ell=120$, and the error exceeds 1\% for
$\ell\geq 1208$. The factor $\mathcal{S}^{(-)}_{\ell\ell'L}$ is defined as
%
\begin{align}
\mathcal{S}^{(-)}_{\ell\ell'L}
\equiv \sqrt{\frac{(2\ell+1)(2\ell'+1)(2L+1)}{16\pi}}
\left[-\ell(\ell+1)+\ell'(\ell'+1)+L(L+1)\right]
\begin{pmatrix}
 \ell & \ell' & L \\
2 & -2 & 0
\end{pmatrix}.
\end{align}
%
Note that $\mathcal{S}_{\ell\ell'L}^{(-)}$ is zero unless $\ell+\ell'+L$
is odd. Finally, the third term in Eq.~(\ref{eq:noise}), $N_\ell$, is
the instrumental noise multiplied by the effect of beam smearing with a
width of $\sigma_b$.
Here we assume that $N_\ell$ is white noise
given by \cite{Katayama:2011eh}
%
\begin{equation}
 N_\ell = \left(\frac{\pi}{10800}\frac{w_p^{-1/2}}{\mu{\rm
	   K\;arcmin}}\right)^2\mu{\rm K}^2\;{\rm str}.
\end{equation}
%
In the actual observations, $N_\ell$ depends on $\ell$ because of, e.g.,
$1/f$ noise and residual foreground emission.
The foreground contribution can be included partially by increasing
$N_\ell$ from the instrumental noise level. The $\ell$-dependent
foreground residual can be incorporated by following, e.g., Appendix C
of Ref.~\cite{Thorne:2017jft}; however, we shall ignore the $\ell$-dependent noise in
this paper. 

We truncate the summation at $\ell_{\rm max}=500$, as the primordial
B-mode decays at $\ell\gtrsim 80$ and noise and lensing B-mode dominate at large
$\ell$. We have confirmed that the main results are not sensitive to 
$\ell_{\rm max}$ as long as we have $\ell_{\rm max}>100$.

\begin{figure}[!ht]
 \includegraphics[width=8cm]{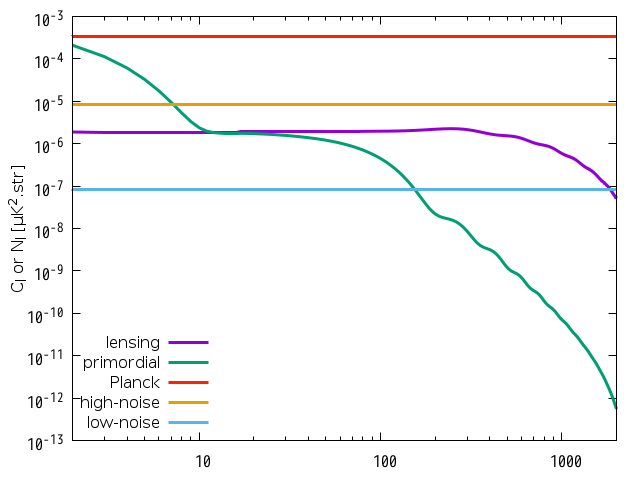}
 \caption{Noise sources assumed in Eq.~(\ref{eq:noise}) where we
 consider the cosmic variance (green), lensing effect of scalar
 perturbations (purple) and the white noise with
 $w_p^{-1/2}=1$ (cyan), $10$ (orange) and $63.1~\mu{\rm K}\cdot{\rm
 arcmin}$ (red), dubbed as ``low-noise'', ``high-noise'' and ``Planck''
 noise models, respectively.}
 \label{fig:noise}
\end{figure}

In this paper, we assume a $0.5$ degree FWHM beam (e.g., LiteBIRD \cite{Matsumura:2013aja}), 
$\sigma_b=0.5\pi/180\sqrt{8\ln 2}=3.7\times 10^{-3}$.
We define three noise models;
(a) a low-noise model with $(w_p^{-1/2},\lambda)=(1~\mu{\rm K}\cdot{\rm arcmin},1)$,
(b) a high-noise model with $(w_p^{-1/2},\lambda)=(10~\mu{\rm K}\cdot{\rm arcmin},1)$,
and (c) a delensed model with $(w_p^{-1/2},\lambda)=(1~\mu{\rm
K}\cdot{\rm arcmin},0)$.
As the lensed B-mode power spectrum at $\ell\ll 10^3$ is approximately
the same as that of white noise with $5~\mu{\rm K}\cdot{\rm arcmin}$
\cite{Lewis:2006fu}, the variance at high multipoles for the case (a) is
dominated by lensing, whereas that for the case (b) is dominated by
noise. The case (c) is nearly an ideal case with complete delensing,
which would be unrealistic but should serve as a useful reference.
The amplitudes of each noise source in Eq.~(\ref{eq:noise}) are shown in 
Fig.~\ref{fig:noise}.

Inverse of the Fisher matrix gives a covariance matrix of the
reconstructed tensor power spectra. The diagonal elements give
$1\sigma$ uncertainties of $\delta\PP_i$ at each bin,
%
\begin{equation}
 \sigma_{\delta\PP_i}^2 = (F^{-1})_{ii}, \label{eq:sigma}.
\end{equation}
%

\section{Results}
\label{sec:results}
\begin{figure}[!ht]
\subfloat[``low-noise'' : $(w_p^{-1/2},\lambda)=(1~\mu{\rm K}\cdot{\rm arcmin},1)$]{
   \includegraphics[width=8cm]{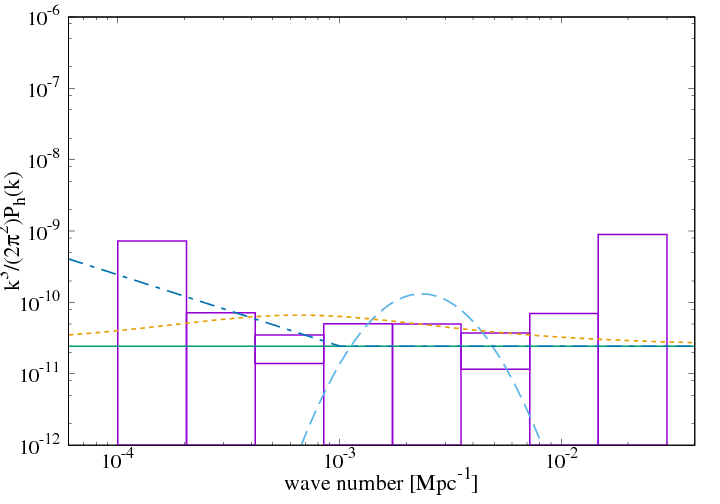}
}
\subfloat[``high-noise'' : $(w_p^{-1/2},\lambda)=(10~\mu{\rm K}\cdot{\rm arcmin},1)$]{
  \includegraphics[width=8cm]{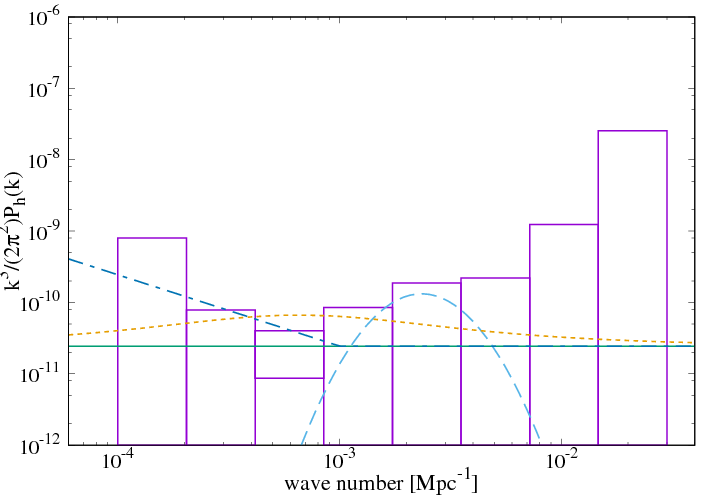}
}
\\
\subfloat[``delensed'' : $(w_p^{-1/2},\lambda)=(1~\mu{\rm K}\cdot{\rm arcmin},0)$]{
   \includegraphics[width=8cm]{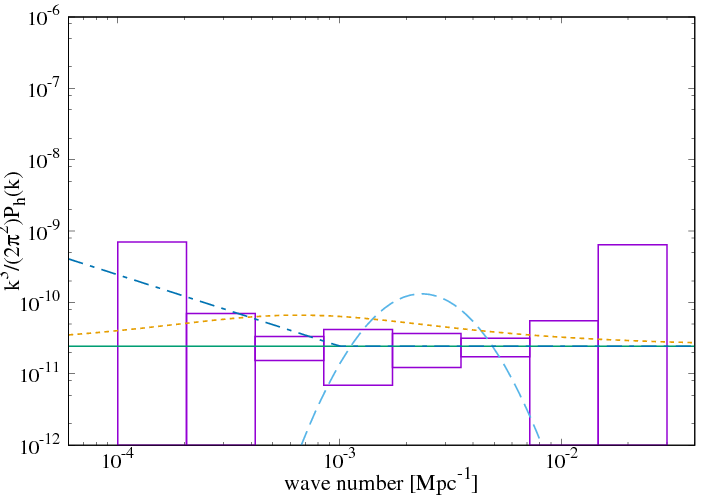}
}
\subfloat[``Planck'' : $(w_p^{-1/2},\lambda)=(63.1~\mu{\rm K}\cdot{\rm arcmin},1)$]{
   \includegraphics[width=8cm]{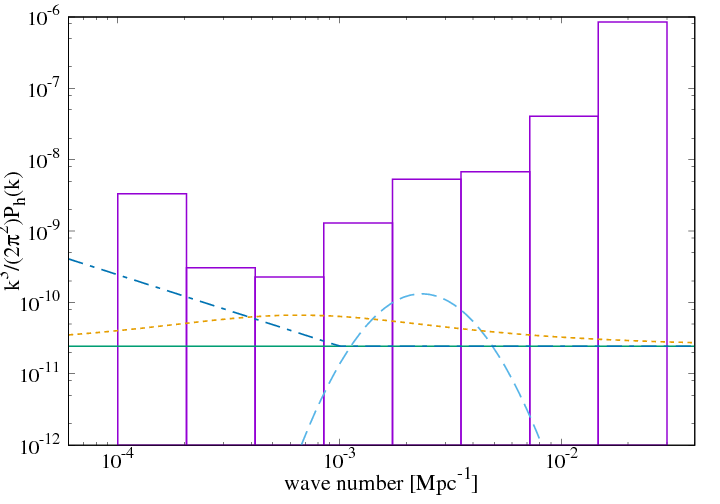}
}
 \caption{Uncertainty of the reconstructed tensor power spectrum from
 B-mode observations. The fiducial model has $r=0.01$ and $n_T=0$, and
 the reconstruction parameters are $k_0=10^{-4}~{\rm Mpc}^{-1}$,
 $k_{N}=3\times 10^{-2}~{\rm Mpc}^{-1}$, and $N=8$. (Top left) Low noise case. (Top
 right) High noise case. (Bottom left) Low noise with complete
 delensing. (Bottom right) Planck noise case.
 The solid line shows the fiducial spectrum, the dashed line an example
 spectrum from the SU(2)-axion model with  $r_*=0.05$, $k_p=2.0\times
 10^{-3}~{\rm Mpc}^{-1}$, and $\sigma=0.4$ \cite{Dimastrogiovanni:2016fuu}, the
 dotted line a massive gravity inflation model with
 $(\alpha,\beta,N_*,T_{\rm reh})=(0.7,1.0,47,10^{10}~{\rm GeV})$,
 and the dot-dashed line a red-tilted spectrum on large scales with
 $(k_1,n_{T1})=(10^{-3}~{\rm Mpc}^{-1},-1.0)$.}
 \label{fig:test1}
\end{figure}

Throughout this paper, we set $f_{\rm sky}=1$.
In Fig.~\ref{fig:test1} we show $\sigma_{\delta\PP_i}$
(Eq.~\ref{eq:sigma}) for
$(r,n_T,k_0,k_N,N,\alpha)=(0.01, 0, 10^{-4}~{\rm
Mpc}^{-1},3\times10^{-2}~{\rm Mpc}^{-1},8,2.04)$.
The solid line shows the fiducial spectrum 
$\PP^{\rm fid}_h$. Each box shows the $1\sigma$ region around the
fiducial spectrum. On large scales, the
uncertainty is mainly due to the cosmic variance. On small scales the
contributions from noise and lensing dominate. 
The covariance matrix including off-diagonal terms is given in
Table~\ref{tab:cov}.

\begin{table}[!ht]
 \centering
 \begin{tabular}{|c||rrrrrrrr|}
\hline
&
\multicolumn{1}{c}{$\delta\PP_1$} & 
\multicolumn{1}{c}{$\delta\PP_2$} & 
\multicolumn{1}{c}{$\delta\PP_3$} & 
\multicolumn{1}{c}{$\delta\PP_4$} & 
\multicolumn{1}{c}{$\delta\PP_5$} & 
\multicolumn{1}{c}{$\delta\PP_6$} & 
\multicolumn{1}{c}{$\delta\PP_7$} & 
\multicolumn{1}{c|}{$\delta\PP_8$} \\
\hline 
$\delta\PP_{1}$ & \fbox{$4.9\times10^{-19}$ } &  $-3.0\times10^{-20}$  &  $2.9\times10^{-21}$  &  $-2.7\times10^{-21}$
&  $2.6\times10^{-22}$  &  $-2.8\times10^{-23}$  &  $5.2\times10^{-23}$  &  $-8.8\times10^{-22}$  \\
$\delta\PP_{2}$ & $-3.0\times10^{-20}$  & \fbox{ $2.2\times10^{-21}$ } &  $-2.5\times10^{-22}$  &  $2.5\times10^{-22}$  
&  $-2.6\times10^{-23}$  &  $3.1\times10^{-24}$  &  $-5.6\times10^{-24}$  &  $9.5\times10^{-23}$  \\
$\delta\PP_{3}$ & $2.9\times10^{-21}$  &  $-2.5\times10^{-22}$  & \fbox{ $1.1\times10^{-22}$ } &  $-1.7\times10^{-22}$  
&  $2.7\times10^{-23}$  &  $-3.5\times10^{-24}$  &  $6.2\times10^{-24}$  &  $-1.0\times10^{-22}$  \\
$\delta\PP_{4}$ & $-2.7\times10^{-21}$  &  $2.5\times10^{-22}$  &  $-1.7\times10^{-22}$  &  \fbox{$6.7\times10^{-22}$}
&  $-1.8\times10^{-22}$  &  $2.5\times10^{-23}$  &  $-4.2\times10^{-23}$  &  $6.9\times10^{-22}$  \\
$\delta\PP_{5}$ & $2.6\times10^{-22}$  &  $-2.6\times10^{-23}$  &  $2.7\times10^{-23}$  &  $-1.8\times10^{-22}$  
& \fbox{ $6.5\times10^{-22}$ } &  $-1.8\times10^{-22}$  &  $2.8\times10^{-22}$  &  $-4.3\times10^{-21}$  \\
$\delta\PP_{6}$ & $-2.8\times10^{-23}$  &  $3.1\times10^{-24}$  &  $-3.5\times10^{-24}$  &  $2.5\times10^{-23}$  
&  $-1.8\times10^{-22}$  & \fbox{ $1.7\times10^{-22}$ } &  $-4.8\times10^{-22}$  &  $8.4\times10^{-21}$  \\
$\delta\PP_{7}$ & $5.2\times10^{-23}$  &  $-5.6\times10^{-24}$  &  $6.2\times10^{-24}$  &  $-4.2\times10^{-23}$  
&  $2.8\times10^{-22}$  &  $-4.8\times10^{-22}$  & \fbox{ $2.1\times10^{-21}$ } &  $-3.9\times10^{-20}$  \\
$\delta\PP_{8}$ & $-8.8\times10^{-22}$  &  $9.5\times10^{-23}$  &  $-1.0\times10^{-22}$  &  $6.9\times10^{-22}$  
&  $-4.3\times10^{-21}$  &  $8.4\times10^{-21}$  &  $-3.9\times10^{-20}$  &  \fbox{$7.5\times10^{-19}$}  \\
\hline
 \end{tabular}
 \caption{Covariance matrix $(F^{-1})_{ij}$ for ``low-noise'' model with $N=8$ and $f_{\rm sky}=1$. 
 The value enclosed in the boxes are the diagonal elements. The wavenumber of each bin is given by $k_n=\alpha^nk_0$ where $\alpha=(k_N/k_0)^{1/N}=2.04$, and see Eq.~(\ref{eq:bandpower}). One can obtain the covariance matrix 
  with $f_{\rm sky}<1$ by multiplying all the elements by $1/f_{\rm sky}$.}
 \label{tab:cov}
\end{table}

We find that the tensor power spectra are best
reconstructed at two wavenumber bins around $k\approx 6\times 10^{-4}$
and $5\times 10^{-3}~{\rm Mpc}^{-1}$. While the precise wavenumbers at
which the spectra are best constrained depend on the choice of bin
sizes, we can understand these values analytically. The B-mode power
spectrum of CMB polarization has two characteristic scales: the
so-called ``reionization bump'' at $\ell\lesssim 6$ and the
``recombinatiom bump'' at $\ell\approx 80$. The wavenumber that gives
the former is $k_{\rm reion}\approx 3/[r_L-r(z_{\rm reion})]$
\cite{Zaldarriaga:1996ke}, where $r_L=14~{\rm Gpc}$ and
$r(z_{\rm reion})\approx 9~{\rm Gpc}$ are the comoving distances to the
surface of last scatter and the epoch of reionization, e.g., $z_{\rm
reion}\approx 8$. We thus obtain $k_{\rm reion}\approx 6\times
10^{-4}~{\rm Mpc}^{-1}$. The wavenumber that gives the latter is
$k_{\rm recomb}\approx 80/r_L\approx 6\times 10^{-3}~{\rm Mpc}^{-1}$.

Usually, the $1\sigma$ regions shrink as we go to higher wavenumbers
where the number of modes is greater; however, we find in
Fig.~\ref{fig:test1} an unusual feature that the $1\sigma$ regions
shrink first, increase at $k\approx 10^{-3}~{\rm Mpc}^{-1}$, and shrink
again at $k\gtrsim 2\times 10^{-3}~{\rm Mpc}^{-1}$. This is due to a gap
(i.e., lack of the signal) between the reionization and recombination
bumps. The transfer function leaves only a small B-mode signal here,
making reconstruction of the initial tensor power spectrum noisy. With
these we understand all the features in Fig.~\ref{fig:test1}.

Can we distinguish between various models of the source of tensor modes
from inflation? In Fig.~\ref{fig:test1} we show some theoretical
predictions of the tensor power spectrum from an SU(2)-axion model with
$(r_*,k_p,\sigma)=(0.05, 2.0\times 10^{-3}~{\rm Mpc}^{-1},0.4)$, from a
massive gravity inflation model with 
$(\alpha,\beta,T_R,g_{*S},N_*,n_{T*})=(0.7,1.0,10^{10}{\rm GeV},100,47,0)$
(see Appendix~\ref{appsub:model}), as well as from a red-tilted spectrum
on large scales with $\PP(k)=(k/k_1)^{n_{T1}}$ for $k<k_1$ and
$\PP_h(k)=\PP_h^{\rm fid}$ for $k\leq k_1$, which resembles predictions
of an open inflation model associated with
a bubble nucleation \cite{Yamauchi:2011qq}. As an example, we show the
spectrum with $(k_1,n_{T1})=(10^{-3}~{\rm Mpc}^{-1},-1)$.
We emphasize that these parameter choices are not at all
robust predictions of the models, but serve only as examples.

To quantify how well we can distinguish models, we calculate the
$\chi^2$ statistic including the off-diagonal elements of the full
covariance matrix. To this end we calculate $\chi^2$ as
%
\begin{equation}
\chi^2 = 
\sum_{i\leq j}^N
\left[\PP_h^{\rm fid}(k_i) - \PP_h^{\rm model}(k_i)\right]
F_{ij}
\left[\PP_h^{\rm fid}(k_j) - \PP_h^{\rm model}(k_j)\right],
\label{eq:chi2}
\end{equation}
%
and the probability to exceed (PTE) defined as
%
\begin{equation}
 P(\chi^2>a,N) = \int_a^\infty\! P(\chi^2,N)\,d\chi^2.
\end{equation}
%
Here, $P(x,n)$ is the $\chi^2$ distribution function for $n$ degrees of freedom,
%
\begin{equation}
 P(x,N) = \frac{1}{2^{N/2}\Gamma(N/2)}x^{N/2-1}e^{-x/2}.
\end{equation}
%
The PTE provides the probability to confuse the theoretically-predicted models
mentioned above with the fiducial power spectrum. For simplicity, we fix the
theoretical model parameters and do not include them in the degrees of
freedom. 

The values of $\chi^2$ and PTE with $N=8$ are 
tabulated in Table \ref{tab:chi2}. For reference, we also compute them for
the Planck observation with the corresponding white noise, 
$w_p^{-1/2}=63.1 \mu {\rm K}\cdot {\rm arcmin}$, which is obtained by averaging
the noise bandpowers in 70, 100, and 148~GHz \cite{Planck:2006aa}.
In the last row in Table \ref{tab:chi2}, we also show $\chi^2$ for
the null hypothesis, which is calculated by setting $\PP_h^{\rm model}(k_i)=0$ in Eq.~(\ref{eq:chi2}).
We find that Planck cannot detect the fiducial spectrum, and furthermore cannot distinguish the three theoretical predictions
from it, since  $\chi^2$ is of order unity and the
corresponding PTE is also unity. On the other hand, 
the future observations with $w_p^{-1/2}=1~\mu$K$\cdot$arcmin can
distinguish SU(2)-axion model and the massive gravity inflation
model with high statistical significance, whereas the open
inflation model is distinguished with moderate significance
because of the cosmic variance at small wavenumbers.

One may be surprised that we can distinguish the models despite the fact
that the error bars appear larger than the differences between some
models and the fiducial spectrum in Fig.~\ref{fig:test1}. This is due to
large correlations between the bins (see Table~\ref{tab:cov}). Indeed, ignoring the off-diagonal elements, i.e.,
$\sigma^2 = \sum_{i}^N F_{ii}\left[\PP_h^{\rm fid}(k_i) - \PP_h^{\rm
model}(k_i)\right]^2$, we find that, for $N\geq 8$ bins, $\sigma^2\ll
\chi^2$. We also find that the values of $\sigma^2$ depend sensitively
on the number of bins used, whereas those of $\chi^2$ with off-diagonal
terms do not. Only when the size of the bins is sufficiently large (see
$N=4$ in Table~\ref{tab:chi2_sig2}) $\chi^2$ and $\sigma^2$ agree because the
bin-to-bin correlation would be suppressed in this case; thus, including
the off-diagonal elements is essential.

So far, we have fixed the cosmological parameters. How would varying
them change our results? Varying $\Omega_M$ and $H_0$ changes the
distance to the last-scattering surface, shifting the B-mode power
spectrum in the $\ell$ space. This would change the relationship between
$k$ and $\ell$, shifting features in the reconstructed tensor power
spectra in the $k$ space. Varying the optical depth $\tau$ changes the
height of the reionization bump, which would affect the amplitude of the
reconstructed power at $k=k_{\rm reion}\approx 6\times 10^{-4}~{\rm
Mpc}^{-1}$. However, in the era when we can make precise measurements of
the B-mode power spectrum, these parameters will be determined so
precisely that their impacts would not be the dominant uncertainty in
the reconstructed power spectra. 

We have also fixed our fiducial tensor power spectrum at a power-law
power spectrum with $n_T\approx 0$. This is because this spectrum is
motivated by single-field slow-roll inflation models, and detecting
difference from it would be a major discovery. Of course, we are free to
use any spectra as the fiducial power spectrum.

\begin{table}[!ht]
\centering
\begin{tabular}{|l|cc|cc|cc|cc|}
\hline
            & \multicolumn{2}{c|}{low-noise} 
            & \multicolumn{2}{c|}{high-noise} 
            & \multicolumn{2}{c|}{delensed}
            & \multicolumn{2}{c|}{Planck} \\\hline
            & $\chi^2$ & PTE
            & $\chi^2$ & PTE
            & $\chi^2$ & PTE
            & $\chi^2$ & PTE  \\\hline
SU(2)-axion & $1.5\times 10^{2}$  & $9.3\times 10^{-28}$
            & $1.2\times 10^{1}$  & $1.4\times 10^{-1}$
            & $6.3\times 10^{2}$  & $1.6\times 10^{-131}$
            & $4.9\times 10^{-1}$ & $1.0$ \\
Massive     & $1.2\times 10^{2}$ & $1.4\times 10^{-21}$
            & $3.4\times 10^{1}$ & $3.3\times 10^{-5}$
            & $3.6\times 10^{2}$ & $7.7\times 10^{-73}$
            & $1.3$              & $1.0$ \\
Red-tilted  & $2.0\times 10^{1}$ & $1.1\times 10^{-2}$
            & $1.6\times 10^{1}$ & $4.3\times 10^{-2}$
            & $2.1\times 10^{1}$ & $6.0\times 10^{-3}$
            & $1.7$              & $9.9\times 10^{-1}$\\
Null hypothesis
                 & $3.1\times 10^{2}$  & $1.1\times 10^{-61}$
                 & $2.1\times 10^{1}$  & $7.0\times 10^{-3}$
                 & $1.4\times 10^{3}$  & $9.0\times 10^{-295}$
                 & $5.7\times 10^{-1}$ & $1.0$ \\ \hline
\end{tabular} 
 \caption{$\chi^2$ and probability to exceed (PTE) for various noise models, 'low-noise', 'high-noise',
 'delensed' and 'Planck', which corresponds to $(w_p^{-1/2},\lambda)=(1.0,1.0)$,
 $(10.0,1.0)$, $(1.0,0.0)$ and $(63.1,1.0)$, respectively.}
 \label{tab:chi2}
\end{table}

\begin{table}[!ht]
\begin{tabular}{|l|cccc||cccc|}
\hline
&\multicolumn{4}{c||}{$\chi^2$} &\multicolumn{4}{c|}{$\sigma^2$} \\\hline
& $N=4$ & $N=8$ & $N=12$ & $N=16$ & $N=4$ & $N=8$ & $N=12$ & $N=16$ \\ \hline
SU(2)-axion & 2.2$\times 10^{3}$ & $1.5\times 10^{2}$ & $2.8\times 10^{2}$ & $2.4\times 10^{2}$ & $7.2\times 10^{2}$ & $2.4\times 10^{1}$ & $3.5\times 10^{1}$ & $1.6\times 10^{1}$ \\
Massive     & 1.4$\times 10^{2}$ & $1.2\times 10^{2}$ & $1.1\times 10^{2}$ & $1.0\times 10^{2}$ & $6.5\times 10^{1}$ & $2.1\times 10^{1}$ & $7.1$ & $1.5$\\
Red-tilted  & 3.0$\times 10^{1}$ & $2.0\times 10^{1}$ & $1.6\times 10^{1}$ & $1.5\times 10^{1}$ & $2.5\times 10^{1}$ & $4.2             $ & $5.5\times 10^{-1}$ & $1.9\times 10^{-2}$ \\
Null hypothesis & 3.2$\times 10^{2}$ & $3.1\times 10^{2}$ & $2.8\times 10^{2}$ & $2.7\times 10^{2}$ & $7.5\times 10^{1}$ & $1.1\times 10^{1}$ & $4.6$ & $1.9$ \\\hline
\end{tabular} 
 \caption{Dependence of $\chi^2$ and $\sigma^2$ on the number of
 bins for ``low-noise'' model.}
 \label{tab:chi2_sig2}
\end{table}

\section{Conclusion}
\label{sec:conclusion}
Reconstruction of the initial tensor power spectrum is complementary to
the usual approach of forward-modeling (i.e., to calculate the B-mode CMB power spectrum from
a given initial tensor power spectrum) because we can test various
models of the early universe directly at the initial power spectrum level,
without having to run Boltzmann solvers. In this paper we have
calculated the covariance matrix of the reconstructed tensor power
spectra in bins of wavenumbers. The $\chi^2$ statistic (Eq.~\ref{eq:chi2}) computed with this covariance matrix (given in Table~\ref{tab:cov} for
the fiducial power spectrum with $r=0.01$ and $n_T=0$ and $1~\mu{\rm
K}\cdot{\rm arcmin}$ noise) can be used to distinguish the tensor power
spectra of one's favorite early universe models from a power-law
power spectrum. We find that reconstructed power spectra in bins of
wavenumbers are highly correlated and thus including the off-diagonal
elements in $\chi^2$ is essential in obtaining the correct answer.

We have tested our algorithm for three models, SU(2)-axion model \cite{Dimastrogiovanni:2016fuu}, massive
gravity inflation (Sec.~\ref{appsub:model}), and open inflation \cite{Yamauchi:2011qq}, and find that future observations
of CMB polarization by, e.g., LiteBIRD \cite{Matsumura:2013aja}, should
be able to distinguish the theoretical predictions of SU(2)-axion,
open inflation, and massive gravity inflation models from a
scale-invariant tensor power spectrum, depending on the model parameters. While we did not perform comprehensive parameter search for
various models in this paper, we developed an interactive web tool to calculate $\chi^2$ for
any parameter values specified by users.  This application is available
on-line at {\tt http://numerus.sakura.ne.jp/research/open/srec/srec.php}. We describe
this tool in Appendix~\ref{appsub:users}. The web tool returns the
covariance matrix, the $\chi^2$ values and the PTE, and draws figures such as
Fig.~\ref{fig:test1}. 

\begin{acknowledgments}
 This work was initiated at the 1st annual symposium of the Innovative Area ``Why Does the Universe Accelerate? -- Exhaustive Study and Challenges for the Future --'' held at the High Energy Accelerator Research Organization (KEK) on March 8-10 in 2017, and was completed at the symposium of the Yukawa International Seminar (YKIS2018a) ``General Relativity -- The Next Generation --'' held at Yukawa Institute for Theoretical Physics in Kyoto University on February 19-23 in 2018.
 This work was supported in part by JSPS KAKENHI Grant Number
 JP16H01098 (T.~H.), JP15H05896 (E.~K.), JP15H05891 (M.~H.), and
 JP15H05888 (M.~S.). T.~H. was also supported by
MEXT-Supported Program for the Strategic Research Foundation at Private Universities,
2014-2018 (S1411024).
\end{acknowledgments}

\appendix
\section{Massive gravity inflation}
\label{appsub:model}
We consider the inflationary massive gravity theory
with the mass term that depends on dynamics of inflation.
%
\begin{align}
m_g^2=f(\phi,\dot\phi)\,.
\end{align}
%
Depending on the form of the function $f$, it may vary 
substantially during inflation. 

The equation-of-motion for the tensor perturbation takes the form,
%
\begin{align}
\ddot\gamma+3H\dot\gamma+\left(\frac{k^2}{a^2}+m_g^2\right)\gamma=0\,.
\end{align}
%
Assuming a very small slow-roll parameter $\epsilon=-\dot H/H^2$, we obtain
%
\begin{align}
\frac{d^2\gamma}{dn^2}+3\frac{d\gamma}{dn}+\left(\frac{k^2}{a^2H^2}+
\frac{m_g^2}{H^2}\right)\gamma=0\,,
\end{align}
%
where $dn=Hdt$. We set $n=n_f$ at the end of inflation.
On superhorizon scales, assuming $m_g^2/H^2\ll1$, the above equation is solved
to give the amplitude at the end of inflation as
%
\begin{align}
\gamma_{k}(n_f)=\gamma_k(n_k)
\exp\left[-\int_{n_k}^{n_f}\frac{m_g^2}{3H^2}dn\right]\,,
\end{align}
%
where $n_k$ is the time at which the mode crosses the horizon,
$k^2/a^2=H^2$, the rms amplitude of which is 
$\langle\gamma_k^2(n_k)\rangle\propto H^2$ as usual.
Thus the spectrum at the end of inflation is given by
\begin{align}
P_T(k;n_f)
&\propto
\exp\left[-\int_{n_k}^{n_f}\frac{2m_g^2}{3H^2}dn\right]\,,
\end{align}
where $n_f-n_k=\ln(k_f/k)$ and $k_f=a(n_f)H$.

Now let us assume the time dependence of $m_g^2$ as
%
\begin{align}
\frac{2m_g^2}{3H^2}=n_{T*}+\beta\alpha\frac{\sinh\alpha n}{\cosh^2\alpha n}\,,
\end{align}
%
where we assume $\alpha\lesssim 1$ but $\beta$ is arbitrary.
We can then easily integrate it to find
%
\begin{align}
\exp\left[-\int_{n_k}^{n_f}\frac{m_g^2}{3H^2}dn\right]
=\exp\left[-n_{T*}N_k+\frac{\beta}{\cosh\alpha(N_k-N_*)}
-\frac{\beta}{\cosh\alpha N_*}\right]\,,
\end{align}
%
where $N_k=n_f-n_k=-\ln(k/k_f)$ is the number of e-folds counted backward
from the end of inflation, and $N_*=n_f$ is the time at which
the feature in the spectrum appears. Since we assumed $\alpha\lesssim1$
and we want $N_*$ to be fairly large $N_*\gtrsim 40-50$ to have an 
observable feature,
 the last term in the exponent is completely negligible.
Thus we obtain
%
\begin{align}
P_T(k;n_f)\propto
\exp\left[-n_{T*}N_k+\frac{\beta}{\cosh\alpha(N_k-N_*)}\right]
=\left(\frac{k}{k_f}\right)^{n_{T*}}
\exp\left[\frac{\beta}{\cosh\alpha(N_k-N_*)}\right]\,.
\end{align}
%
Thus the spectrum is the product of a power-law component
and a factor peaked at $N=N_*$. The enhancement factor is
$e^\beta$ relative to the baseline.

\section{User's manual of \appname}
\label{appsub:users} 
We developed a web tool,
\appname\footnote{http://numerus.sakura.ne.jp/research/open/srec/srec.php}, to
compute the Fisher matrix of reconstructed initial tensor power
spectra. In this section, we provide a brief instruction of this tool.

\appname assumes the cosmological parameters given in
Table \ref{tab:param}. It returns a Fisher matrix and
a covariance matrix, and makes a plot of 
the fiducial power spectrum of tensor perturbations with error bars
where the fiducial spectrum is assumed to be a power-low given in
Eq.~(\ref{eq:fiducial}).

The covariance matrix is then used to compute $\chi^2$ and the PTE for
various early universe models. Three kinds of model power spectra
that are introduced in the main text are provided in the tool as
`built-in models'. One can also upload numerical data of a power
spectrum as `custom model'.

In the main page of the tool, we define the parameters controlling the
Fisher analysis and plots, which are categorized into four tabs: `Basic',
`Drawing', `Built-in models' and `Custom models'. One can get
information on each parameter in these tabs when one hovers over
parameter names.
In `Basic' tab, one can specify the amplitude and the spectral index of
the fiducial spectrum, the number of bins, and noise sources. In
`Drawing' tab, one can adjust the vertical and horizontal axes of the
plot as well as the scale (logarithmic or linear). 
In `Built-in models', one can set the
model parameters of SU$(2)$-axion, open inflation, and the massive
gravity models that are introduced in the main text, and also select 
the presence or absence of each model spectrum in the plot.
Finally, in `Custom' models, one can upload favorite power spectrum data
in a simple text format.

After setting the parameters, clicking the `MAKE PLOT' button generates
a plot in the PNG format. If one selects the presence of
some model spectra, the corresponding $\chi^2$'s and PTE's are also
tabulated below the plot. The Fisher and covariance matrices are
provided in the text format at the link below the plot.
This text file contains four blocks; the first two blocks are the Fisher
matrices with and without the cosmic variance, and the remainings are
the corresponding covariance matrices. The parameters and results
including the uploaded spectrum, if exists, are preserved for a few days
on the system. 

Note that specifications and appearance of our web tool are subjected to
change without prior notice for improvement.

\end{document}